# Graded Index Microstructured Polymer Optical Fibers for Terahertz Applications

Tian Ma[1,2], Andrey Markov[1], Lili Wang[2], Maksim Skorobogatiy[1]

[1]École Polytechnique de Montréal, Montreal, Québec H3C3A7, Canada
[2]Xi'an Institute of Optics and Precision Mechanics of CAS, Xi'an, Shannxi 710119, China

**Abstract:** Graded index microstructured polymer optical fiber incorporating a specially designed air-hole array featuring variable air-hole diameters and inter-hole separation is proposed, fabricated and characterized in view of the fiber potential applications in low-loss, low-dispersion terahertz guidance. The proposed fiber features simultaneously low chromatic and intermodal dispersions, as well as low loss in the terahertz spectral range. We then experimentally demonstrate that proposed fibers exhibit smaller pulse distortion, larger bandwidth and more reliable excitation when compared to the porous fibers of comparable geometry.

## 1. Introduction

The terahertz frequency range has large potential for various technological and scientific applications, such as sensing, imaging, communications and spectroscopy. Terahertz sources are generally bulky and designing efficient THz waveguides for flexible delivery of the broadband THz radiation would be a big step towards simplification of alignment of THz systems for these applications. Many point THz devices, such as sources, filters, sensor cells, detectors can be connected together in a single system in order to simplify the system integration. Usage of THz fibers is crucial for various applications, for example, endoscopy and crevice inspection.

The main complexity in designing terahertz waveguides is the fact that almost all materials are highly absorbent (over ~ 1m propagation) in the terahertz region. Since the lowest absorption loss occurs in dry gases, an efficient waveguide design typically maximizes the fraction of power guided in the gas. Typically, low absorption transmission is achieved using porous or hollow core THz fibers. Previous works [1-6] showed that introducing porosity in the core is an efficient methodology for low-loss terahertz waveguide structural design. It was also theoretically and experimentally demonstrated that the introduction of porosity enables broadening of the main transmission window compared to a non-porous fiber with the same diameter, and also blue-shifting of the transmission peak to higher frequencies [1].Large pores size makes it convenient to fill the fiber with an analyte, this possibility has been explored for bacteria concentration detection in a wide range [3].

Waveguide might exhibit a significant variation of its optical properties as a function of frequency (refractive index, group velocities, etc.), thus leading to pulse broadening, and hence, signal amplitude reduction. Therefore waveguide dispersion management is an important issue when guiding broadband pulses. Dispersion managed THz fibers have been proposed [7, 8]. These dispersion managed fibers are usually single mode and, hence, only optimized for individual mode dispersion. However, when guiding large bandwidth pulses (like a typical TDS-THz pulse), the intermodal dispersion is a more significant issue.

When designing a dielectric terahertz fiber, one must face a trade-off between dispersion and loss in dielectric fibers. The frequency band can be roughly divided into 3 zones (see Fig. 1). In the first one, the losses and dispersion are low but the mode is completely delocalized, which makes the usage of this frequency region absolutely impractical. Propagation is single mode in the second zone, losses are higher compared to the first frequency region but still low, however the mode quickly changes its localization, thus leading to very high dispersion. In the third region refractive indices are almost constant, hence, dispersion is low, however the losses are high in this zone and also a fiber becomes multimode. For efficient guiding of very large bandwidth THz pulses, fiber design should be optimized to manage losses and dispersion both in the second (single mode) and third (multi-mode) zones.

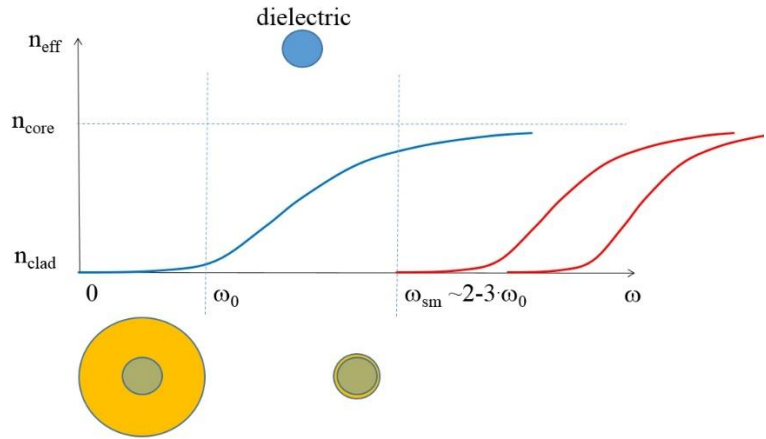

Fig. 1. Dispersion/loss trade-off in the dielectric THz Waveguides

Multimode fibers (MMF) have been widely used for communication systems [9-12]. However, intersymbol interference caused by the intermodal dispersion, which arises because of the diverse group velocities of different modes, limits the achievable bit rate-distance products [13]. Thus, control of intermodal dispersion becomes a critical problem for communication systems based on optical fibers.

Since 1990s, graded index polymer optical fibers (GI-POF) have been thoroughly investigated in view of their short length telecommunication applications [14-20]. Recently, Y. Akimoto et al. [14] reported the fabrication and experimental investigation of the doped PSt-based GI-POF. The graded refractive index distribution was achieved by controlled radial dopant concentration. The authors confirmed that this fiber has a high bandwidth (4.4 GHz at 655 nm with fiber length of 50 m) and low attention (166-193 dB/km at 670-680 nm) and can be used for home networks. Previously, R. Lwin et al. [20] reported another type of graded index fiber with graded index profile obtained by porous cladding with specially designed air hole positions and diameters, namely graded index microstructured polymer optical fiber (GI-mPOF). They demonstrated experimentally that the bandwidths of the proposed GI-mPOFs were much wider than those of the commercial GI-POF.

With the development of terahertz communication systems, such intermodal dispersion managed fibers are also demanded. However there are reasons preventing direct scaling of these fibers into the THz range. In the visible or near infrared the fibers are designed for a very narrow frequency range, whereas THz pulses usually have very broad frequency ranges. The size of the graded index fibers for visible and infrared is in hundreds of wavelengths they are designed for, and obviously scaling of these fibers to THz will be impractical. Also, graded index fibers for visible and infrared guide mostly in glass or plastics, whereas THz radiation is highly absorbed by most materials.

In this paper, we propose a novel porous fiber which is specifically designed for practical applications in terahertz range, such as terahertz communication and imaging, which demand low intermodal dispersion and low loss simultaneously. The proposed fiber design is a combination of the graded index fibers used in visible range and porous fibers for terahertz frequency range. It features an air-hole array with inconsistent air-hole diameters and lattice constants, which is used to create a radially graded index distribution. We confirm that, compared to a traditional microstructured polymer optical fiber with uniform air-hole diameters and lattice constants, the proposed fiber structure reduces the intermodal dispersion, while keeping the losses at a low level. To the best of our knowledge, this is the first time such an intermodal dispersion managed fiber are reported for terahertz applications.

## 2. Fiber structural design and fabrication

In order to obtain a refractive index profile similar to that of the conventional GI-POF shown in Ref. [14], we used a hexagonal lattice of 5 rings of air-holes with gradually varied diameters and lattice constants to replace the doped polymers. The azimuthal average refractive index was approximated by a power law form as shown in Eq. (1).

$$n(r) = n_0 (1 - 2a(r/R)^g)^{1/2} \quad (1)$$

where $a$ is refractive index difference between the fiber material and the cladding, $R$ is the radius of this fiber, $g=2$. Here we assume that the localized refractive index of microstructured porous fiber is determined by the local air filling fraction $\eta$, which could be defined as

$$n(r) = \eta \cdot n_{air} + (1-\eta) \cdot n_{polymer} \quad (2)$$

Based on these two equations, we calculated the air-hole diameters and hole to hole distances for the parameter g in Eq. (1) equal to 2.

All fibers presented in this paper were fabricated using commercial rods of low density polyethylene (LDPE) known as one of the lowest absorptive polymers in the THz region [21]. The fiber preforms of GI-mPOFs were fabricated using drilling method. The air-hole array with designed structural parameters were drilled on the transversal surface of a LDPE rods with 1.5 inch outer diameter and 12 cm length. Then the fabricated preforms were drawn down to fibers with outer diameter (OD) of 1.3-1.5 mm in a fiber drawing tower. In order to make a reference measurement, a microstructured optical fiber with uniform diameters and lattice constants was also fabricated using the same material and fabrication technique. The structural parameters of this reference fiber match the average of the corresponding parameters of the GI-mPOF. Fig. 2 shows the cross section of the two fibers.

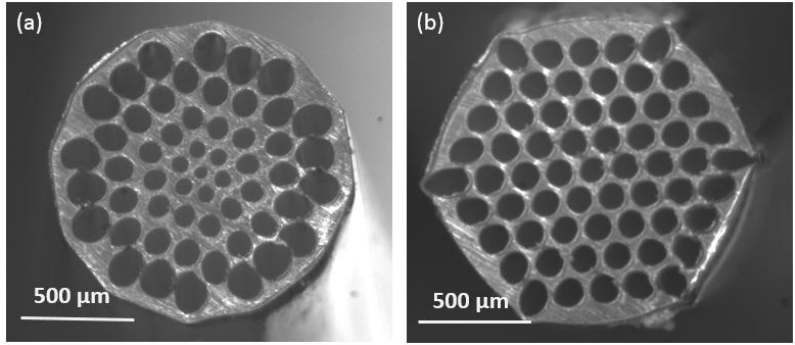

Fig. 2. The cross-sections of (a) GI-mPOF (OD=1.35mm) and (b) mPOF (OD=1.47mm)

## 3. Numerical model and simulation results.

To investigate the optically properties of the proposed GI-mPOF numerically, we used the commercial software COMSOL to solve for the complex effective index and field distribution of the guided modes. Fig. 3 shows the calculated modal refractive index ($n_{mod}$) and group velocities ($v_g$) of the guided modes of the two fibers. In these figures, the dots colors represent the logarithmic flux coupling coefficient ($C_m^2$), and here we only show the modes with excitation efficiency higher than 0.01. . In the group velocity versus frequency graph, we notice that the variation of the values of group velocities is smaller for the case of the GI-mPOF.

Correspondingly, the variable $C_m$ refers to the normalized amplitude coupling coefficient computed from the overlap integral of the respective flux distribution of the m-th mode with that of the 2D Gaussian beam of the source. Specifically, the definition of $C_m$ is based on the continuity of the transverse field components across the input interface (i.e. cross-section of the fibers) between the incident beam and the excited fiber modes [22]:

$$C_m = \frac{1}{4} \int dxdy (\mathbf{E}_m^*(x,y) \times \mathbf{H}_{input}(x,y) + \mathbf{E}_{input}(x,y) \times \mathbf{H}_m^*(x,y))$$

$$\times \frac{1}{\sqrt{\frac{1}{2}\mathrm{Re}\int dxdy(\mathbf{E}_{input}^*(x,y) \times \mathbf{H}_{input}(x,y))} \times \sqrt{\frac{1}{2}\mathrm{Re}\int dxdy(\mathbf{E}_m^*(x,y) \times \mathbf{H}_m(x,y))}} \quad (3)$$

To model the field structure of the source, we assume a y-polarized 2D Gaussian beam whose fields are normalized to carry power P, then limited by an aperture of radius R as follows:

$$\vec{E}_{Input}(x,y) = \vec{x} \cdot \sqrt{\frac{2P}{\pi\sigma^2}} \cdot \exp\left[-\frac{y^2}{2\sigma^2}\right]$$

$$\vec{H}_{Input}(x,y) = \vec{y} \cdot \frac{1}{\sqrt{\mu_0/\varepsilon_0}} \cdot \sqrt{\frac{2P}{\pi\sigma^2}} \cdot \exp\left[-\frac{y^2}{2\sigma^2}\right] \quad (4)$$

$$for\ x^2 + y^2 \leq R^2;$$

$$\vec{E}_{Input}(x,y) = 0,\ \vec{H}_{Input}(x,y) = 0\ for\ x^2 + y^2 > R^2$$

where the Gaussian beam waist parameter σ is related to the full-width hall-maxima by field as $FWHM = 2\sigma\sqrt{2 \cdot \ln 2}$, $\vec{x}$ and $\vec{y}$ are the unit vectors in x- and y-directions, $\sqrt{\mu_0/\varepsilon_0}$ is the intrinsic impedance of vacuum, $R$ is equal to the radius of the fiber (~0.65 mm). The frequency dependence of the beam waist was measured experimentally and then fitted by a linear function of the input wavelength $\sigma = 0.96\lambda$.

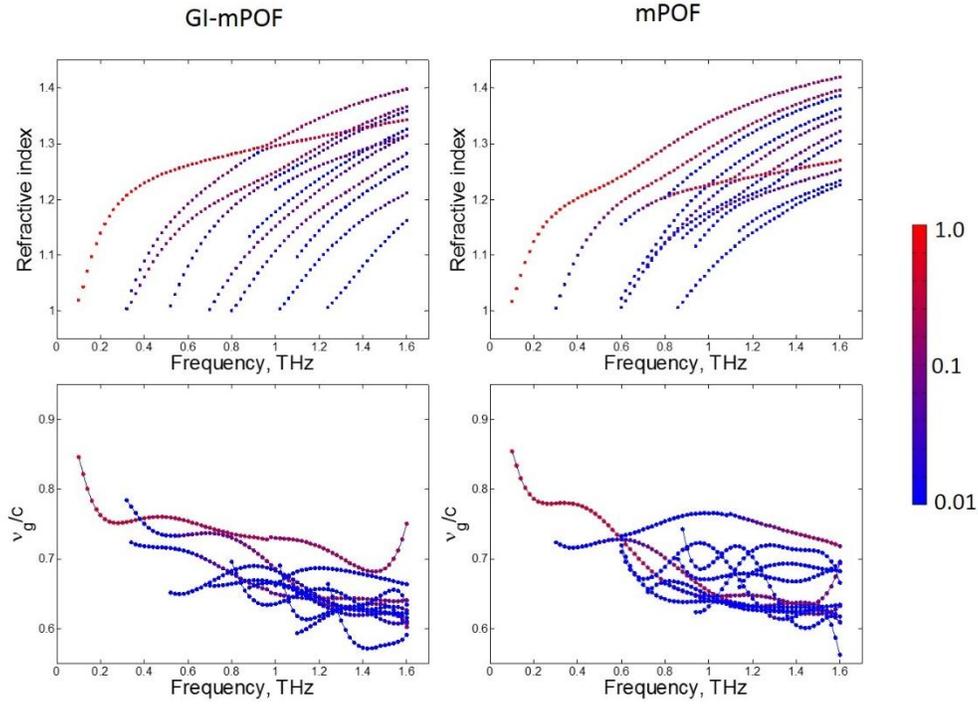

Fig. 3. (a) The modal refractive indices and (b) the group velocities of the proposed GI-mPOF and the traditional mPOF. The dots colors represent to the logarithmic flux coupling coefficient of each modes at the given frequency.

As shown in Fig. 4, the fundamental mode of the GI-mPOF (shown as a black curve) is predominately excited in the entire frequency range, whereas coupling into higher order modes is correspondingly lower as opposed to the fiber with uniform holes. In the case of traditional mPOF, higher order modes have higher coupling coefficients compared to the fundamental mode (black curve) above 0.7 THz.

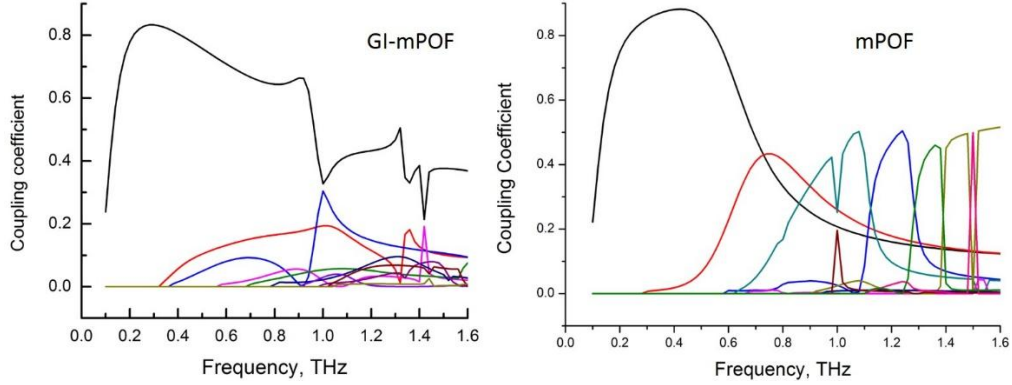

Fig. 4. Coupling efficiency by power for the proposed GI-mPOF and the traditional mPOF

Based on the modal refractive indices and group velocity, we calculated the individual mode (or waveguide) dispersion ($\sqrt{\langle D^2 \rangle}$) and the intermodal dispersion ($\langle \Delta v_g^{-2} \rangle$). The waveguide dispersion was computed based on the first order derivative of the modal propagation constant. The intermodal dispersion was defined as the standard deviation of the modal delay and can be given as [23]:

$$\langle \Delta v_g^{-2} \rangle = \langle v_g^{-2} \rangle - \langle v_g^{-1} \rangle^2 \qquad (5)$$

where $\langle M \rangle$ denotes the average of the variable $M$ and is defined as $\langle M \rangle = \sum_j M_j C_j^2$. Assuming pulse intensity at the fiber input z=0 as $I(t,0) \sim \exp\left(-\left(\frac{2t}{\tau_0}\right)^2\right)$, pulse width after propagation over a fiber length z can be found as follows:

$$\langle \tau^2(z) \rangle = \frac{\tau_0^2}{8} + z^2 \left( \left[\langle v_g^{-2} \rangle - \langle v_g^{-1} \rangle^2\right] + 2\frac{\langle D^2 \rangle}{\tau_0^2} \right) \qquad (6),$$

where one can see the contribution of intermodal and individual mode dispersions into pulse broadening. For the calculations we assume that the initial pulse is 1 ps long.

In Fig. 5, we depict the results of the computation for the two types of dispersion of both the proposed GI-mPOF and traditional mPOF. As we can see from Fig. 5, both the individual mode dispersion and the intermodal dispersion of the GI-mPOF has been reduced significantly compared to the traditional porous fiber with an array of uniform holes. As a result, we expect that the proposed fiber structure will considerably decrease the pulse width broadening by both lowering the impacts of higher order modes and by reducing intermodal dispersion.

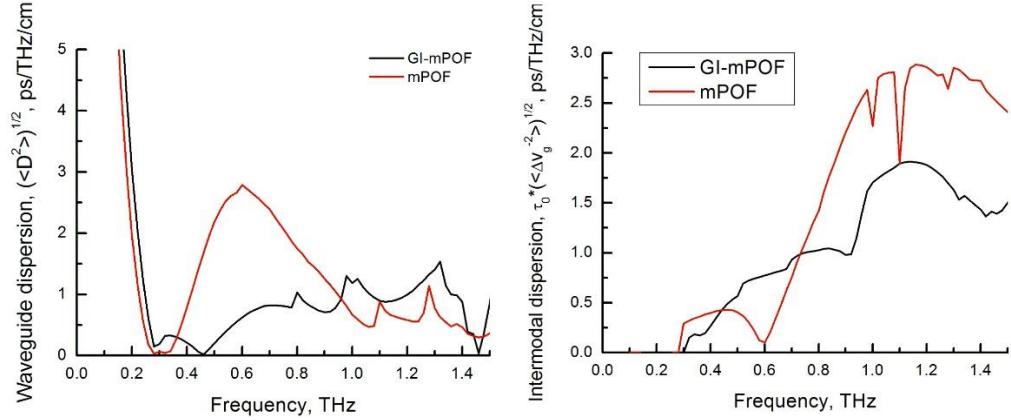

Fig. 5. (a) The individual mode dispersion and (b) the intermodal dispersion of the two fibers. The red solid lines are the dispersion properties of the proposed GI-mPOF, while the black lines show that of the traditional mPOF.

## 4. THz-TDS measurement

All measurements in our experiment were obtained by using a modified terahertz time-domain spectroscopy (THz-TDS) setup. The setup consists of a frequency doubled femtosecond fiber laser (MenloSystems C-fiber laser) used as pump source and two identical GaAs dipole antennae used as THz emitter and detector yielding a spectrum range of 0.1 to 3.0 THz. However, because of the lower dynamic range and increased material losses in the fiber at higher frequencies, we only considered the spectrum range of 0.2 to 1.5 THz in the following sections.

With a parabolic mirror mounted on the translation rails, our setup allows measuring the waveguides up to 45 cm. Fig. 6 illustrates the experimental setup where the fiber was placed between the two parabolic mirrors. To obtain the transmission properties of the fiber, we used the cutback method in the measurement. The input facet of the fiber was fixed by gluing to an aperture, while the output facet was cut in steps. Both the input and output ends of the fiber were placed at the focal points of the parabolic mirrors.

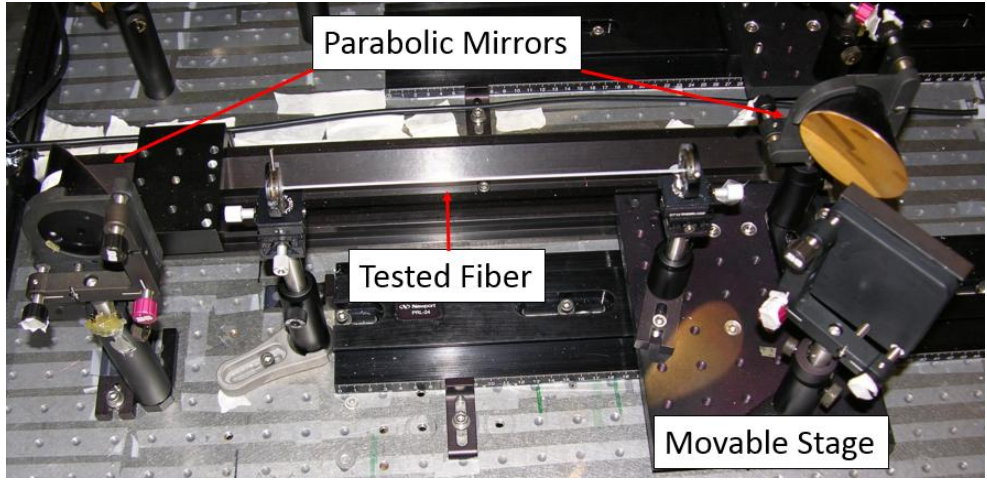

Fig. 6. Experimental setup with the fiber mounted in the apertures.

In Fig. 7, we present three temporal traces of the THz electric field measured at different fiber lengths. As expected based on the simulations, lower intermodal dispersion of the GI-mPOF leads to the reduction of the pulse broadening. With lower difference of the group velocities, most of the modes reach the output end of the fiber in one envelope. Other higher order modes with larger mode delay were also restrained because of the low coupling coefficient. Meanwhile, in the case of traditional mPOF, the output pulse has been divided into several groups because of the higher difference of the modal group velocities.

Another interesting phenomenon can be observed from the temporal traces of the two fibers. The output electric field of the traditional mPOF is much smaller than that of the proposed GI-mPOF. This phenomenon is caused by the weaker mode confinement of the traditional mPOF. As shown by the simulated distribution of the fundamental mode at 0.5 THz (see Fig. 7 (b)), the mode of the proposed GI-mPOF is mostly concentrated in the central region of the fiber. Meanwhile in the case of the traditional mPOF, the fundamental mode leaves the core and propagates mainly on the plastic/air interface. In our experiment, this evanescent field is blocked along with stray light by the metal aperture and partly absorbed by the glue.

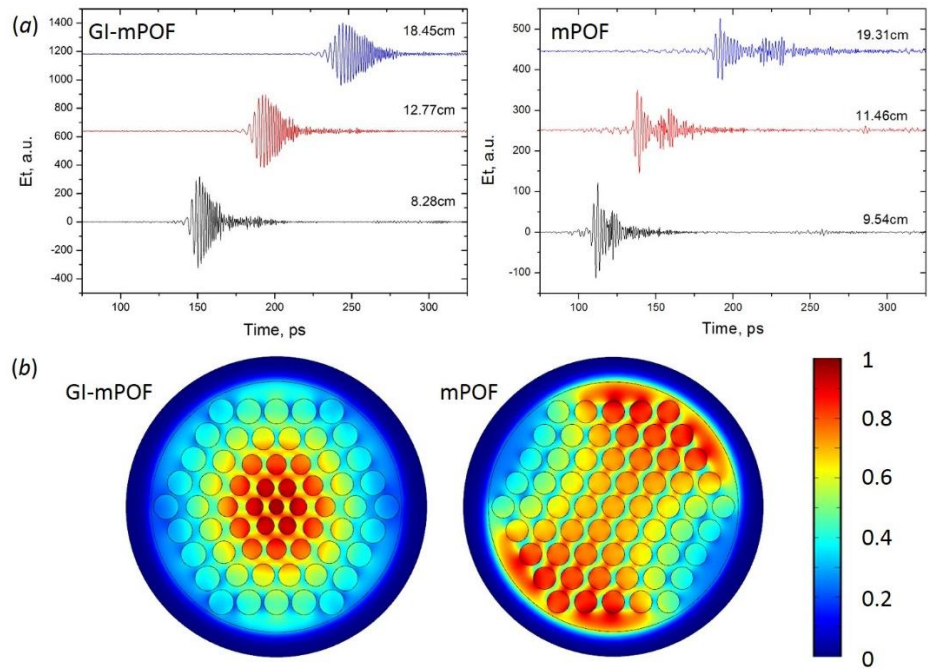

Fig. 7. (a) The time-domain traces of THz electric field measured at different fiber lengths of the proposed GI-mPOF (left) and the traditional mPOF (right). Black trace is the THz field after propagating a short distance in fiber, red trace is for longer distance and blue trace is for the whole fiber. The initial lengths of the test fibers are about 20 cm. (b) Mode profiles simulated at 0.5 THz for these two fibers.

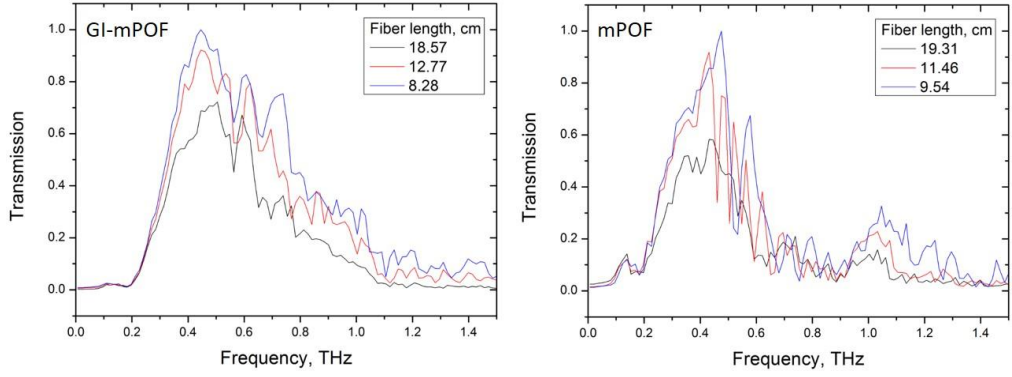

Fig. 8. The transmission spectra of the proposed GI-mPOF (lift) and the traditional mPOF (right).

We also compare the transmission spectra of these two fibers in the frequency domain, as demonstrated in Fig. 8. A wider transmission band from 0.2 to 1.5 THz is observed for the proposed GI-mPOF, whereas it stops at 0.8 THz and has a small peak at 1.1 to 1.2 THz in the case of the traditional mPOF. At the same time, the interaction of the excited modes, which can be shown by ripples in the higher frequency range of the transmission spectra, has been restrained by both lower intermodal dispersion and low coupling coefficients of higher order modes in the case of GI-mPOF. Based on the transmission spectra, we also computed the absorption losses of the fibers. In Fig. 9 we demonstrate The experimentally measured transmission losses of the proposed GI-mPOF (red squares), calculated absorption losses of the fundamental mode of the fiber (black squares), and the bulk absorption losses of low density polyethylene (black line) used in the calculations ($\alpha_{PE}[cm^{-1}] \approx 0.14 \cdot f^2 [THz]$).

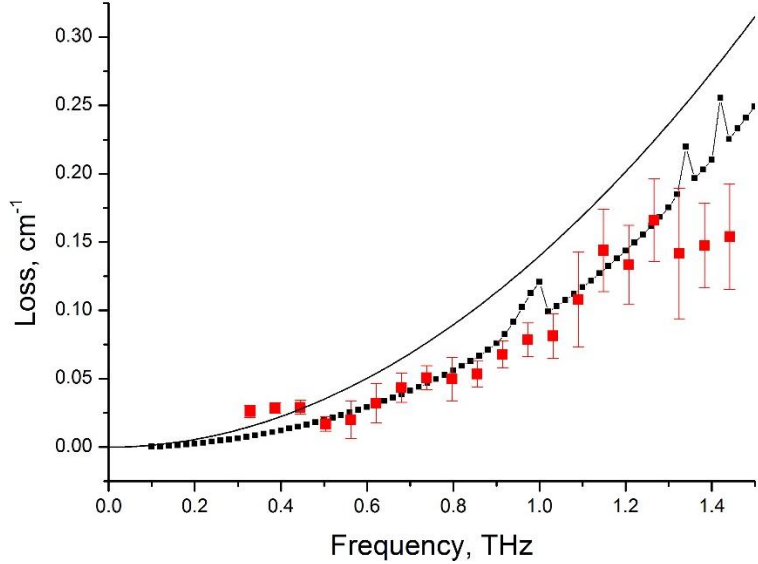

Fig. 9. The experimentally measured transmission losses of the proposed GI-mPOF (red squares), calculated absorption losses of the fundamental mode of the fiber (black squares), and the bulk absorption losses of low density polyethylene (black line).

## 5. Conclusion

A graded index microstructured fiber which is designed for reducing intermodal dispersion has been proposed for applications in THz regime. The radially graded index distribution of the proposed fiber is achieved by an air-hole array with inconsistent air-hole diameters and lattice constants. Large air holes at the outer layers of the proposed GI-mPOF lead to a better modal confinement to the center of the fiber and enhance the output electric field, as compared to the traditional mPOF with uniform air-hole diameters and lattice constants.

In this paper, using the finite element software, we investigated theoretically the modal properties of the proposed GI-mPOF and the traditional mPOF. Simulation results show that the proposed GI-mPOF design suppresses the excitation of higher order modes and reduces the intermodal dispersion. Transmissions of these two fibers were also measured in a THz-TDS using the cut-back method. According to experimental result, the proposed fiber structure improved the output pulses quality as all the modes reach the output facet of the fiber in one time-domain envelope. At the same time, the frequency domain transmission band has been broadened. Good agreement between experimental data and theoretical results has been observed.